\documentclass[pra,aps,twocolumn,superscriptaddress,longbibliography]{revtex4-1}
\usepackage[unicode=true,pdfusetitle,bookmarks=true,bookmarksnumbered=false,bookmarksopen=false,breaklinks=false,pdfborder={0 0 0},backref=false,colorlinks=false]{hyperref}
\hypersetup{colorlinks,linkcolor=myurlcolor,citecolor=myurlcolor,urlcolor=myurlcolor}
\usepackage{braket,colortbl,amsthm,amsmath,cleveref,amssymb,txfonts}\definecolor{myurlcolor}{rgb}{0,0,0.7}
\usepackage{bbm}
\usepackage{graphics,graphicx}
\usepackage{color}
\usepackage{amssymb}
\usepackage{amsthm}
\usepackage{amsfonts}
\usepackage{float}
\usepackage{graphicx}
\usepackage[format = plain,labelfont = bf,up, textfont = normal , up, justification=raggedright, singlelinecheck =false]{caption}
\usepackage{subcaption}
\usepackage{tabularx}
\usepackage{amsmath}
\usepackage{braket}

\usepackage{graphicx}
\usepackage[utf8x]{inputenc}
\usepackage{color,soul}
\usepackage{amsmath}
\usepackage{braket}
\usepackage{latexsym}
\usepackage{amssymb}
\usepackage{amsthm}
\usepackage{bm}
\usepackage{graphics,epstopdf}
\usepackage{color}\usepackage{amsmath}
\usepackage{enumitem}
\usepackage{fmtcount}
\usepackage{booktabs}
\def\Tr{\text{Tr}}

\begin{document}
\title{Noisy quantum batteries}


\author{Kornikar Sen}
\affiliation{Harish-Chandra Research Institute,  A CI of Homi Bhabha National Institute, Chhatnag Road, Jhunsi, Allahabad 211 019, India}

\author{Ujjwal Sen}
\affiliation{Harish-Chandra Research Institute,  A CI of Homi Bhabha National Institute, Chhatnag Road, Jhunsi, Allahabad 211 019, India}

\begin{abstract}
In realistic situations, physical systems can not be completely isolated from its environment. Its inevitable interaction with the environment can influence the working process of the device. In this paper, we consider two-qubit quantum batteries where one qubit of the battery is successively interacting with the spins present in the surrounding environment. We examine the effect of the interaction on the maximum amount of energy that can be extracted from the battery using unitaries. In this context, we use the notion of locally passive states. In particular, we examine the behavior of the amount of extractable work from the noisy battery, initially prepared in a locally passive or ordinary pure state, having a fixed initial entanglement, with the number of interactions the qubit has gone through. We also examine the amount of locally extractable work from the noisy battery. We realize though the amount of extractable energy, be it global or local, as a whole will decrease with the number of spins of environment it interacted with, but if we increase the time interval of the interaction with each spin,  after a cut off value of the interval, the small time behavior shows a peculiarity, 
i.e., the extractable energy within a single interaction starts to increase with time. The cut-off time indicates the Markovian-to-non-Markovian transition of the interaction. We also observe a non-Markovian increase in extractable energy from the Markovian scenario.
\end{abstract}
\maketitle

\section{Introduction}
Quantum technologies are gradually starting to overpower their classical counterparts. Some of the important quantum ingredients fueling the improvement are entanglement \cite{ent1,ent2,ent3}, coherence \cite{coh1, coh2, coh3, coh4}, discord~\cite{dis1, dis2, dis3}, etc. 

An important part of the quantum world containing miniaturized machines is the quantum battery. As far as we know, Alicki and Fannes were the first to discuss quantum batteries~\cite{alicki}. Quantum batteries are described using a quantum mechanical system that is initially in a specific state and a corresponding Hamiltonian that describes the energy of the system. Using a unitary is the conventional way to charge or discharge a battery. After employing unitary operations to extract all available energy, the battery's state becomes passive. 

To investigate the behavior of quantum batteries, several models have been considered, including the Sachdev-Ye-Kitaev~\cite{qb1}, interacting spin chains~\cite{qb3}, Lipkin-Meshkov Glick~\cite{qb2}, the Dicke quantum battery~\cite{qb4,qb5}, non-Hermitian PT-symmetric models~\cite{qb6}, and so on~\cite{qb7,qb8,qb9,qb10,qb11}. A charger is also introduced, which is a quantum mechanical system connected with the battery. The charger can transfer energy from itself to the battery through a global operation on the composite system consisting of the battery and the charger~\cite{ch1,ch2,ch3,qb1,ch5,ch6}. Different unique methods have been introduced for fast charging of the battery, such as by measuring the charger's state~\cite{c3}, using catalysis~\cite{c2}, or using an Otto machine~\cite{c1}. 

Because of the inevitable interaction of a system with its environment, it is obvious that a fully charged battery will not remain in that state for an indefinite period of time. In several works, the effect of dissipation in quantum batteries has been explored~\cite{oq1,oq2,oq3}. In Ref.~\cite{oq}, authors have shown that batteries can be charged faster and can acquire higher maximum energy in presence of Markovian as well as non-Markovian noise [see also~\cite{oq4}]. In Ref.~\cite{oq5}, fundamental bounds on the rate of charging of a noisy battery are provided. 

The extraction of energy from a bipartite quantum battery using local unitary operations is discussed in Ref.~\cite{ref20}. After the extraction of all locally available energy using local unitaries, the battery reaches a locally passive state. In the same work, i.e., Ref.~\cite{ref20}, relations between maximum extractable work using local and global unitary operations are determined as a function of the entanglement shared between the two sub-systems of the battery. Moreover, the maximum globally extractable work from locally passive states is also presented. The authors showed that the maximum globally and locally extractable work from a pure state having fixed entanglement is a monotonically decreasing function of that entanglement, whereas the maximum amount of globally extractable work from a locally passive state is a monotonically increasing function of the same.

In this work, we consider a two-qubit quantum battery where one qubit of the battery is open to an environment which consists of an infinite number of spins. Each of the environment's spins arrives one at a time to interact with the qubit for a small but fixed amount of time, say $\delta t$. At first, we investigate the amount of globally or locally extractable work from the battery after each of the battery's interactions with the environment's spins. 
The extractable work decreases as the number of interacted spins increases.

Because after each time interval, $\delta t$, a new spin of the environment arrives to interact with the battery's qubit, the interaction is Markovian after every time duration $\delta t$. However the behavior of the extractable energy in-between $\delta t$, depends on the length of the interval. The interaction within the time intervals is also Markovian for relatively smaller values of $\delta t$. But if we keep on increasing the value of $\delta t$, after a threshold value, the interaction within the time interval becomes non-Markovian. We investigate the effect of this Markovian and non-Markovian interaction between the bath and the qubit separately by varying $\delta t$. The amount of extractable energy within the time of interaction is a monotonically decreasing function of time for Markovian interaction, but if we choose $\delta t$ large enough that the interaction with each spin becomes non-Markovian, the extractable work begins to increase after a cutoff time within each $\delta t$. It's as if the qubit initially loses energy to the environment during the time interval $\delta t$, but after a while, it begins to absorb energy back from the environment.

Though there are results considering this type of interaction between the battery and the environment's constituents, described by two-level systems ~\cite{col1,col2,col3,col4}, non of them have considered a quantum battery having two constituents. Furthermore, the behavior of the extractable work with entanglement shared between the two constituents in the presence of such noise has, to the best of our knowledge, never been explored.

The rest of the paper is organized as follows: In Sec \ref{sec2}, we briefly recapitulate about the noiseless quantum batteries. The exact model of the battery, the environment, and their interaction are described in Sec \ref{sec3}. In Sec \ref{sec4}, we observe the maximum amount of globally extractable work from two-qubit quantum batteries in the presence of noise, which are initially in an arbitrary but certain locally passive pure state and share a fixed entanglement. In addition, we calculate the maximum amount of globally and locally extractable work from the noisy battery while it is initially prepared in a pure state. We examine the effect of non-Markovianity, by increasing the time interval of interaction of each spin of the environment with the battery, in Sec. \ref{sec5}. Finally, we conclude in Sec. \ref{sec6}.

\section{Isolated Quantum battery}
\label{sec2}
A quantum battery is a quantum mechanical system described by a state, $\rho$, and a Hamiltonian, $H$, both of which act on the same Hilbert space, $\mathcal{H}$. The amount of energy that can be extracted from the battery, using a unitary operation, $U$, is given by
\begin{equation*}
    W=\text{Tr}{\left(\rho H\right)}-\text{Tr}{\left(U\rho U^\dagger H\right)}.
\end{equation*}
The state corresponding to a Hamiltonian, which does not contain any unitarily extractable energy, is called passive. Passive states commute with the Hamiltonian, and if, for a particular permutation of the energy basis, the eigenvalues of the passive state are situated in increasing order, then the corresponding eigenvalues of the Hamiltonian will be in decreasing order, and vice versa. 

The maximum amount of extractable energy, using unitary operation, from a quantum battery, $(\rho,H)$, can be expressed in terms of the corresponding passive state, $\sigma_\rho$, in the following way~\cite{alicki}
\begin{equation}
    W_{max}=\max_U\left[\text{Tr}{\left(\rho H\right)}-\text{Tr}{\left(U\rho U^\dagger H\right)}\right]=\text{Tr}{(\rho H)}-\text{Tr}{(\sigma_\rho H)}. \label{eq3}
\end{equation}
Since $\rho$ is transformed to the passive state, $\sigma_\rho$, using a unitary operation, $\sigma_\rho$ has the same eigenvalues as $\rho$.

Let us now consider a bipartite quantum battery, $(\rho_{12},H_{12})$, defined on a composite Hilbert space $\mathcal{H}_1\otimes\mathcal{H}_2$. The maximum amount of extractable energy from $\rho_{12}$, using only local unitary operations, can be expressed as~\cite{ref20}
\begin{eqnarray*}
     W^l_{max}&=&\max_{U_1\otimes U_2}\left[\Tr{\left(\rho_{12} H_{12}\right)}-\Tr{\left(U_1\otimes U_2\rho_{12} U_1^\dagger\otimes U_2^\dagger H_{12}\right)}\right]\\
     &=& \Tr{(\rho_{12} H_{12})}-\Tr{(\sigma^l_{\rho_{12}} H_{12})}.
\end{eqnarray*}
Here $\sigma^l_{\rho_{12}}$ is the state from which no energy can be extracted using local unitary operations. Thus $\sigma^l_{\rho_{12}}$ is called a locally passive state. 
If we consider a local Hamiltonian, $H_{12}=H_1\otimes I_2+I_1\otimes H_2$, acting on the Hilbert-space $\mathcal{H}_1\otimes \mathcal{H}_2$, where $I_1$ and $I_2$ are identity operators on $\mathcal{H}_1$ and $\mathcal{H}_2$, respectively, then the subsytems, say $\sigma^l_1=\Tr_{2}\left({\sigma^l_{\rho_{12}}}\right)$ and $\sigma^l_2=\Tr_{1}\left({\sigma^l_{\rho_{12}}}\right)$, of the locally passive state, $\sigma^l_{\rho_{12}}$, will commute with, respectively, $H_1$ and $H_2$. Moreover, for a particular permutation of the local energy bases the eigenvalues of $\sigma^l_{1/2}$ will be in opposite order of the eigenvalues of $H_{1/2}$, whereas the eigenvalues of $\sigma^l_1$ ($\sigma^l_2$) is equal to the eigenvalues of $\rho_1=\Tr_2\left(\rho_{12}\right)$ ($\rho_2=\Tr_1\left(\rho_{12}\right)$). Here $\Tr_{1/2}$ denotes taking trace over the Hilbert space $\mathcal{H}_{1/2}$.

\section{model}
\label{sec3}
We consider a two qubit quantum battery described by the following local Hamiltonian
\begin{equation*}
    H_{12}=e_1\sigma^z_1\otimes I+e_2I\otimes \sigma^z_2,\label{eq1}
\end{equation*}
acting on the Hilbert-space $\mathcal{H}_1\otimes \mathcal{H}_2$, where $e_1>e_2$ and $\sigma^z_i$ is the Pauli-$z$ operator acting on the $i$th qubit. $2e_1$ and $2e_2$ are the energy gaps between the two energy levels of the ``first" and ``second" qubit of the battery, respectively. The second qubit is open to interact with a spin bath, $\mathcal{B}$, consisting of spin-$\frac{1}{2}$ particles. $i$th spin of the bath is described by the following Hamiltonian
\begin{equation*}
    H_{B_i}=h\sigma_{B_i}^z,
\end{equation*}
 acting on the single-spin Hilbert-space, $\mathcal{H}_{B_i}$. Here $\sigma^z_{B_i}$ is the Pauli-$z$ matrix acting on the $i$th spin. We denote the joint Hilbert-space of the complete bath by $\mathcal{H}_B=\otimes_i \mathcal{H}_{B_i}$. 

The second qubit of the battery interacts with only one spin of $\mathcal{B}$ for a short interval of time, $\delta t$. After that, the same qubit interacts with another spin of the bath again for same $\delta t$ amount of time. And the process continues. When one particular spin is interacting with the qubit, others remain isolated. The interaction between the second qubit and the $i$th spin is given by
\begin{equation*}
 H_{int}(\delta t)=k(\sigma_2^x\otimes \sigma_{B_i}^x+\sigma_2^y\otimes\sigma_{B_i}^y),
\end{equation*}
where $\sigma^x_2$ ($\sigma^x_{B_i}$) and $\sigma^y_2$ ($\sigma^y_{B_i}$) denote, respectively, Pauli-$x$ and Pauli-$y$ matrices acting on the second qubit ($i$th spin). Here $k$ is a constant which tunes the coupling of the interaction. Moreover, we assume that the spins of $\mathcal{B}$ do not interact with each other. Hence the total Hamiltonian of the composite system consisting of the battery and the bath can be written as
\begin{equation}
    H_{int}=H_{12}\otimes I_B+I_1\otimes H_{int}(\delta t)\otimes I_{B_{\bar{i}}}+I_{12}\otimes_{i} H_{B_i}. \label{eq2}
\end{equation}
Here $I_B$, $I_1$ and $I_{12}$ denote the identity operators acting on $\mathcal{H_B}$, $\mathcal{H}_1$, and $\mathcal{H}_1\otimes \mathcal{H}_2$, respectively. The identity operator acting on all the spins of the bath, except the $i$th spin, is denoted by $I_{B_{\bar{i}}}$. The interactions between the battery and the bath is considered to be Markovian. Thus after the interaction of the $i$th spin of the bath, when the next spin, say $i+1$, comes to interact with the qubit, that $(i+1)$th spin does not have any memory about the previous $i$ interactions.

Let initially all the individual spins of the bath be in the following state
\begin{equation*}
    \rho_{B_i}=\left[\begin{matrix} p_0&0\\0 &p_1 \end{matrix}\right],
\end{equation*}
and initial state of the battery be $\rho_{12}(0)$. Here $p_0$ and $p_1$ are given by $\frac{e^{-\beta h}}{e^{-\beta h}+e^{\beta h}}$ and $\frac{e^{\beta h}}{e^{-\beta h}+e^{\beta h}}$, where $\beta$ is the inverse of temperature.  Within the first time interval, $\delta t$, the second qubit of the battery interacts with one spin of $\mathcal{B}$, lets name it the ``first" spin. Therefore, after time $\delta t$, the state of the battery becomes
\begin{equation*}
    \rho_{12}(\delta t)=\text{Tr}_{B_1}\left[U(\delta t)\rho_{12}(0)\otimes \rho_{B_1}U^\dagger(\delta t)\right],
\end{equation*}
where $U(\delta t)=\text{exp}[-j\delta t(H_{12}\otimes I_{B_1}+I_1\otimes H_{int}(\delta t)+I_{12}\otimes H_{B_1})]$ and $\text{Tr}_{B_i}$ denotes tracing out the $i$th spin. Here we have denoted the complex number, $\sqrt{-1}$, by $j$.
In the next time interval, i.e $(\delta t,2\delta 2]$, the battery interacts with another spin of the bath, say the second spin, which has the same initial state, $\rho_{B_2}=\rho_{B_i}$. After this interaction, the state of the composite system consisting the battery and the bath is
\begin{equation*}
    \rho_{12}(2\delta t)=\text{Tr}_{B_2}\left[U_2(\delta t)\rho_{12}(\delta t)\otimes \rho_{B_2}U_2(\delta t)^\dagger\right].
\end{equation*}
Hence the recurrence relation for $n$ such interactions can be written as
    \begin{equation*}
    \rho_{12}(n\delta t)=\text{Tr}_{B_n}\left[U(\delta t)\rho_S((n-1)\delta t)\otimes \rho_{B_n}U(\delta t)^\dagger\right].
\end{equation*}
 For all the numerical derivations, unless specified, we will take $e_1=2e$, $e_2=e$, $g=1.5e$, $\beta B=10$, $k=e$, and $\delta t=0.2$.

In the following section, we examine the maximum amount of extractable work from the interacting two-qubit quantum battery, initially prepared in a pure state, as a function of time as well as the entanglement shared between the two qubits.
\section{Extractable work from noisy quantum battery}
\label{sec4}
Consider a two-qubit quantum battery, initially in the pure state $\ket{\xi}=c_0\ket{00}+c_1\ket{01}+c_2\ket{10}+c_3\ket{11}$, where $\{\ket{0},\ket{1}\}$ denotes the eigenbasis of $\sigma_1^z$ or $\sigma_2^z$. The logarithmic negativity of the state, which measures the entanglement content within the two qubits, is given by~\cite{LN1,LN2,LN3,LN4}
\begin{equation*}
    E=\log_2(2|c_0c_3-c_1c_2|+1).
\end{equation*}

In the following sub-sections, we consider a quantum battery, initially prepared in a pure state, interacting with a spin bath described using the Hamiltonian given in Eq. \eqref{eq2}. We discuss how much energy is possible to extract from that battery after its interaction with one or several spins of the bath.


\subsection{Global work extraction from locally passive batteries}
\begin{figure*}
\centering
	\includegraphics[scale=0.30]{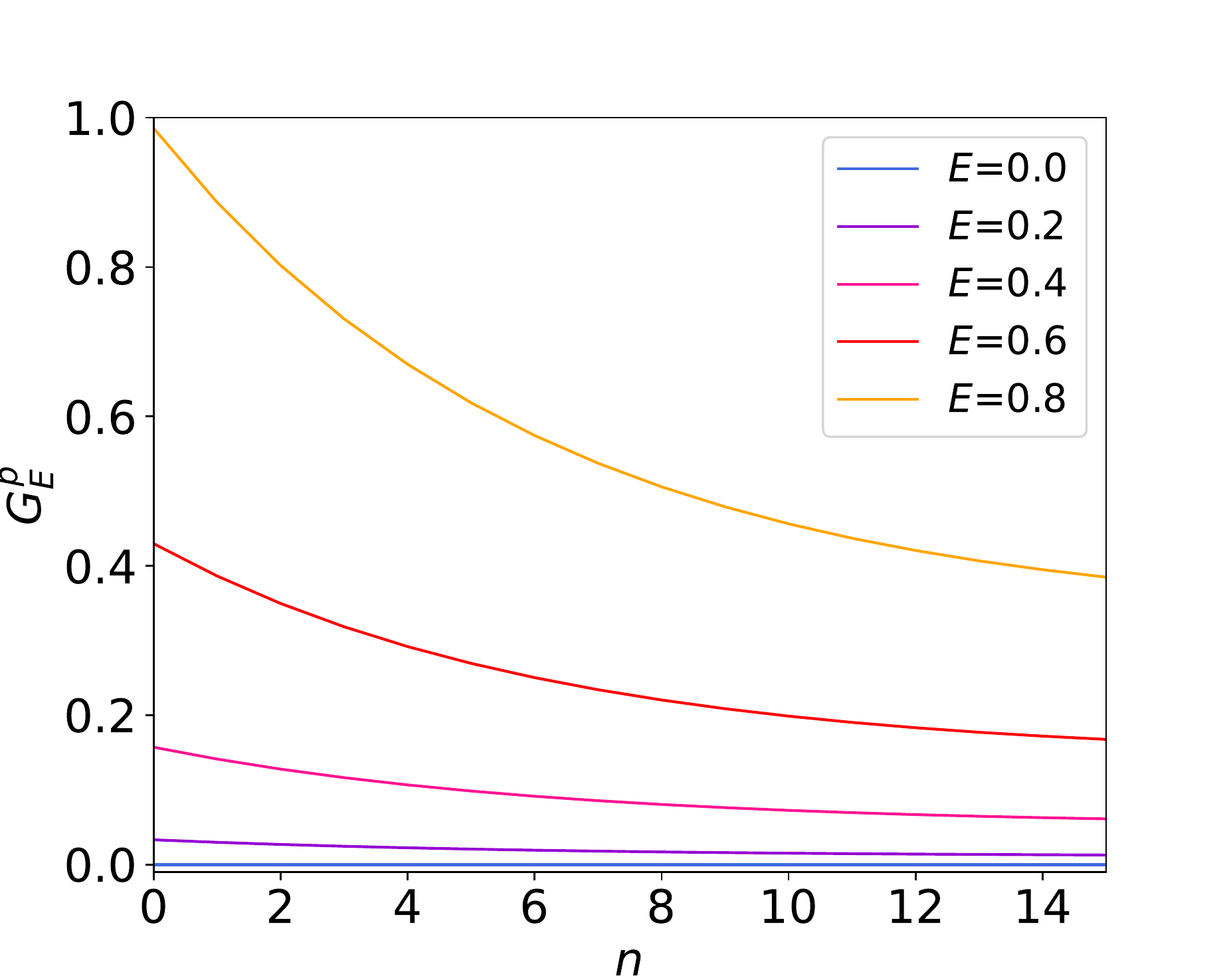}
	\includegraphics[scale=0.30]{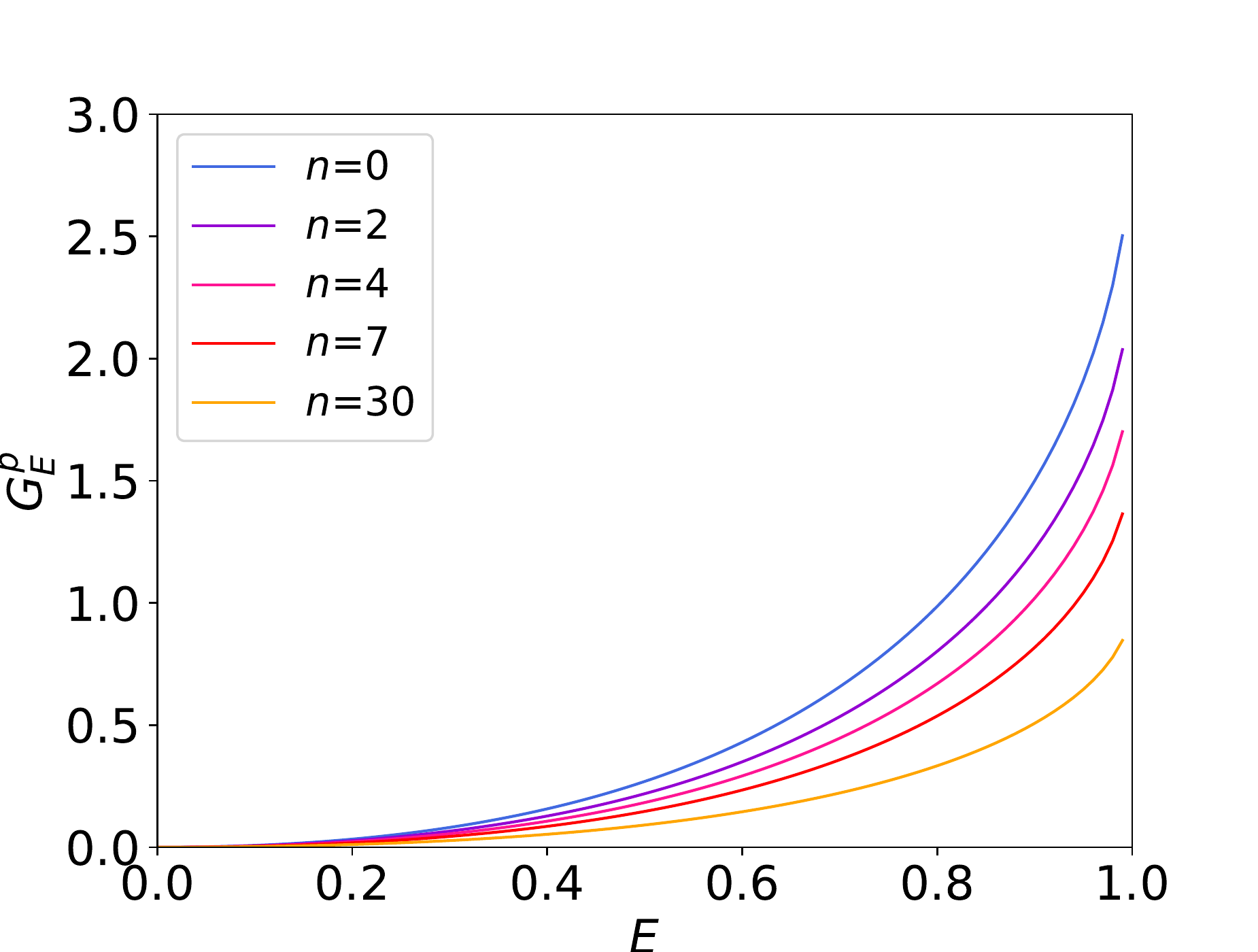}
	\includegraphics[scale=0.30]{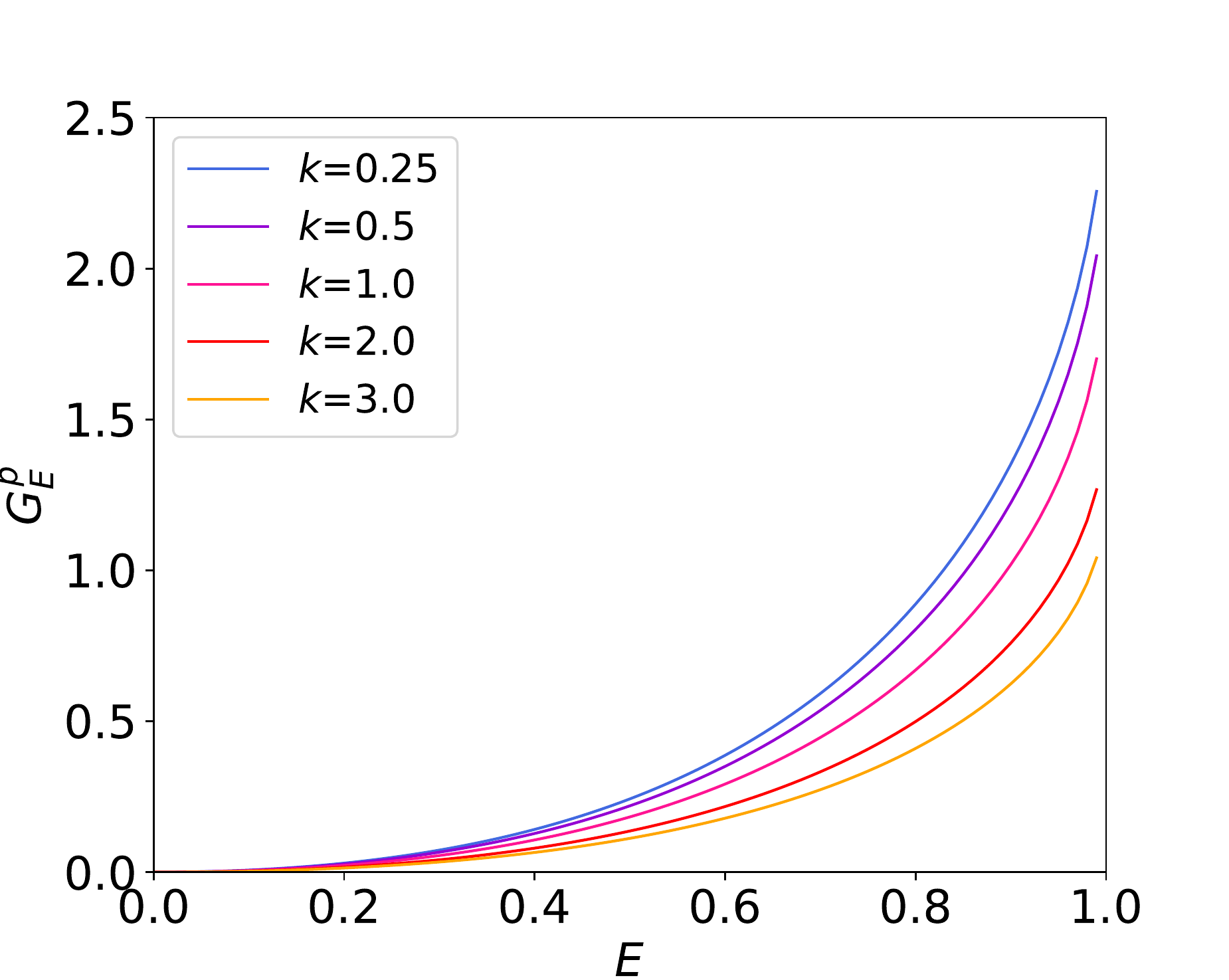}
	\caption{Amount of globally extractable work from an interacting quantum battery, initially prepared in locally passive state. In the left panel, we plot $G_E^p(n)$ as a function of $E$ for different values of entanglement, $E$, viz. $E$=0.0 (blue line), 0.2 (violet line), 0.4 (pink line), 0.6 (red line), 0.8 (yellow line). In the middle and right panels we plot the same as a function of $E$. The fixed values of $n$ considered in the middle panel are 0 (blue line), 2 (violet line), 4 (pink line), 7 (red line), and 30 (yellow line) and the fixed values of $k$ considered in the right panel are 0.25 (blue line), 0.5 (violet line), 1.0 (pink line), 2.0 (red line), and 3.0 (yellow line).  All the vertical axes represent values of $G_E^p(n)$ which is plotted in the units of $e$. The horizontal axes represents two different quantities, $n$ (left panel) and $E$ (middle and right panel), both of which are dimensionless.}
	\label{fig1}
\end{figure*} 

\begin{figure*}
\centering
	\includegraphics[scale=0.30]{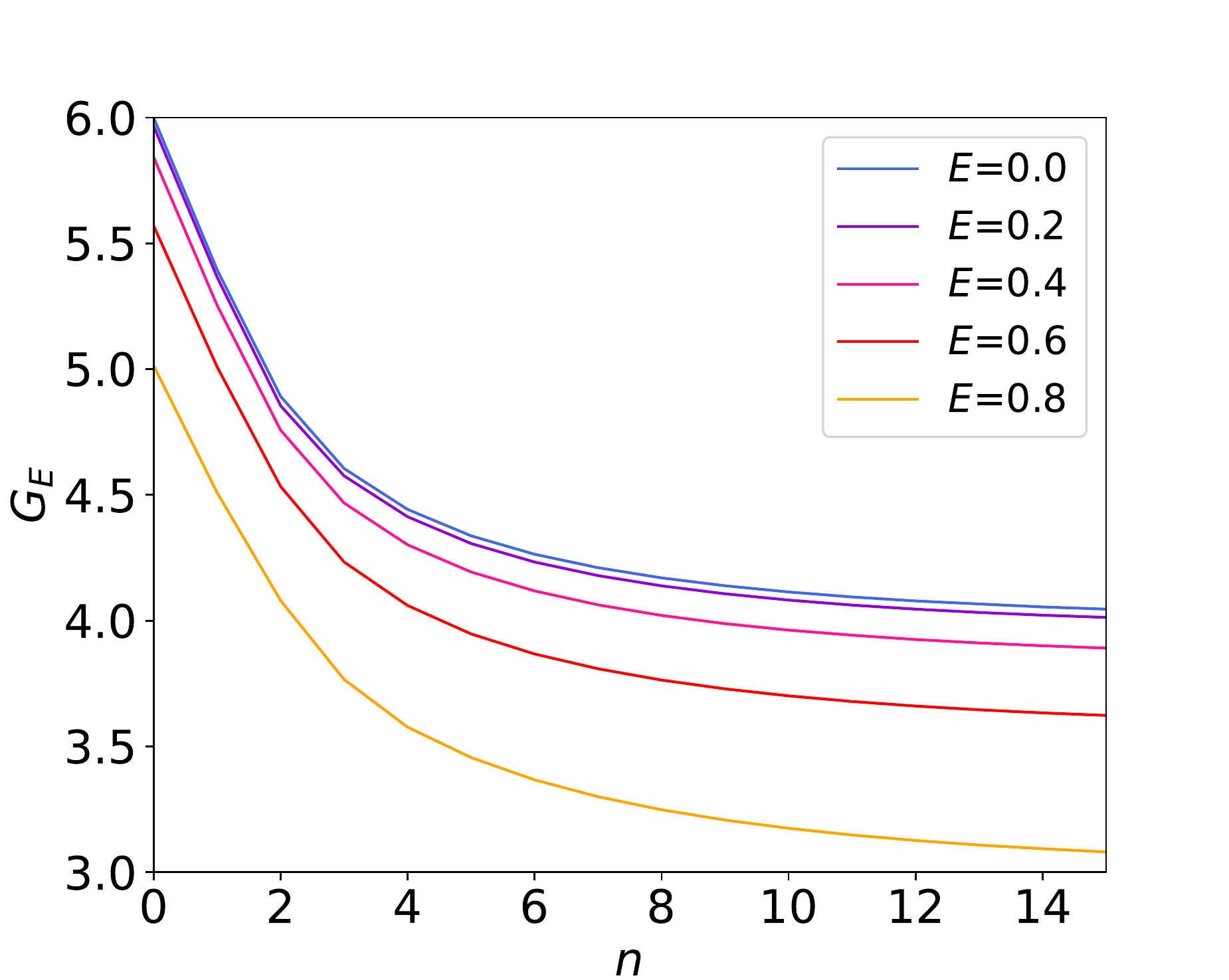}
	\includegraphics[scale=0.30]{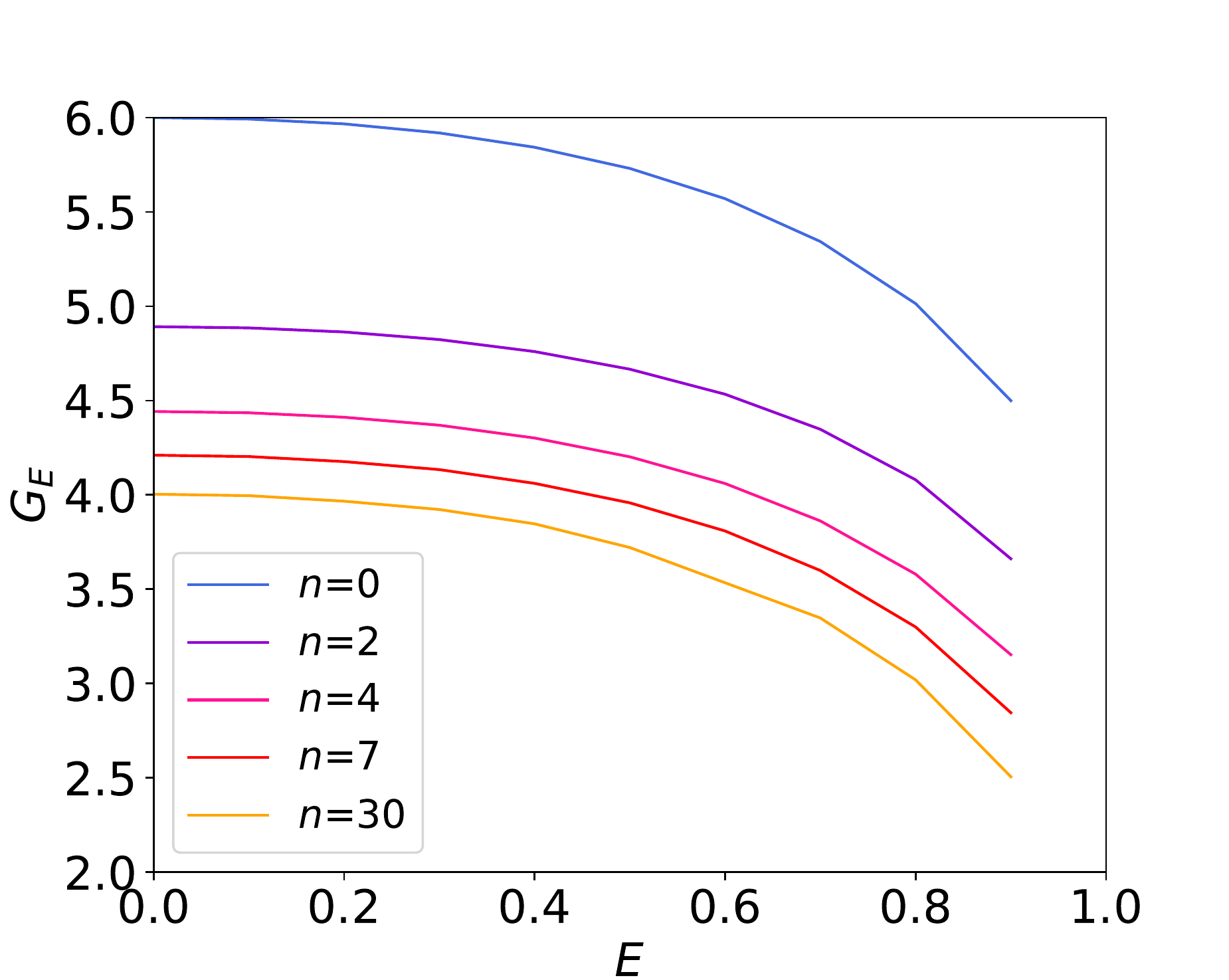}
	\includegraphics[scale=0.30]{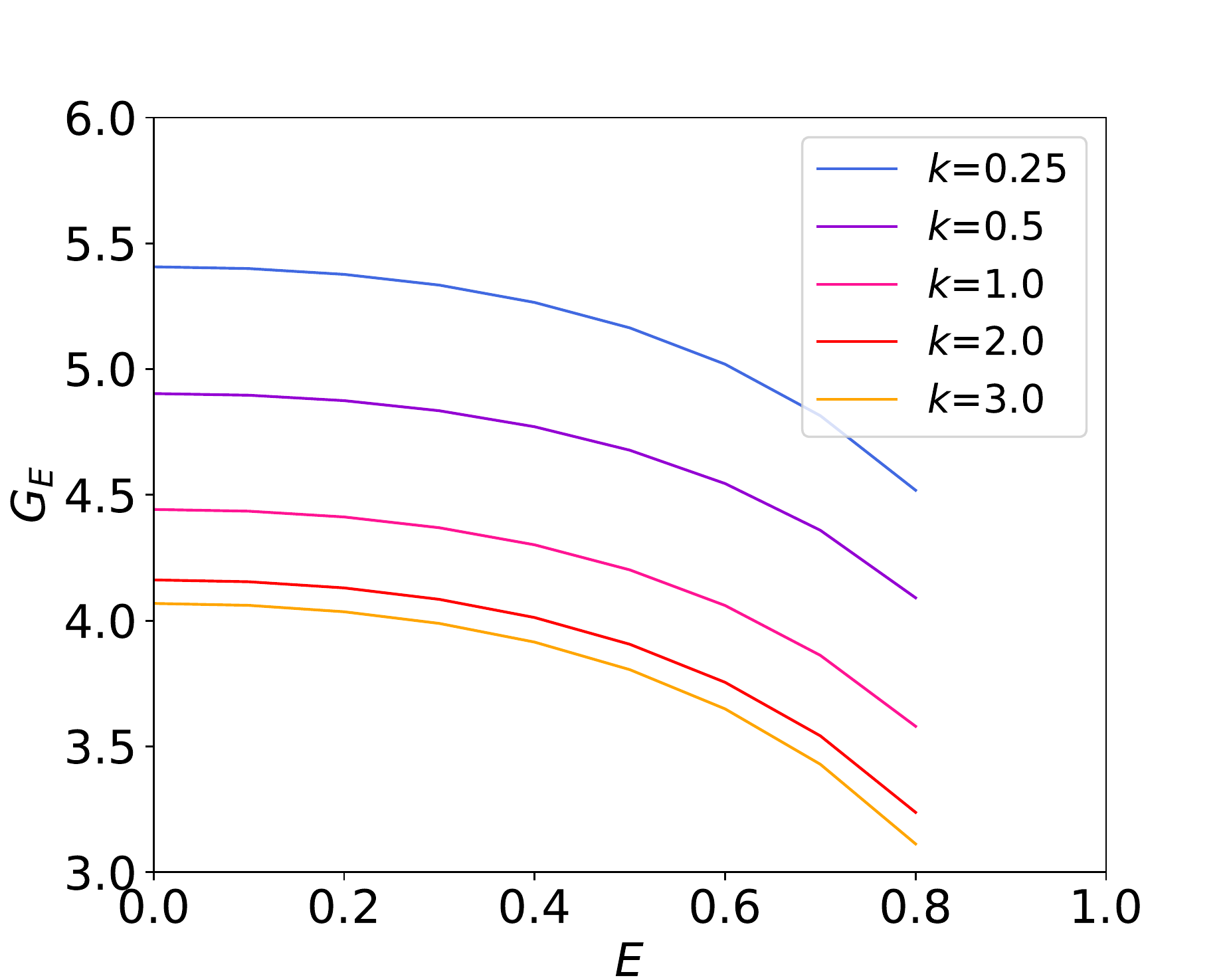}
	\caption{Behavior of extractable work from noisy quantum battery, initially sharing a fixed entanglement. The considerations are same as Fig. \ref{fig1} with the only difference that here the vertical axis represents maximum amount of globally extractable work, maximized over the set of pure initial states with entanglement, $E$.}
	\label{fig2}
\end{figure*} 

\begin{figure*}
\centering
	\includegraphics[scale=0.30]{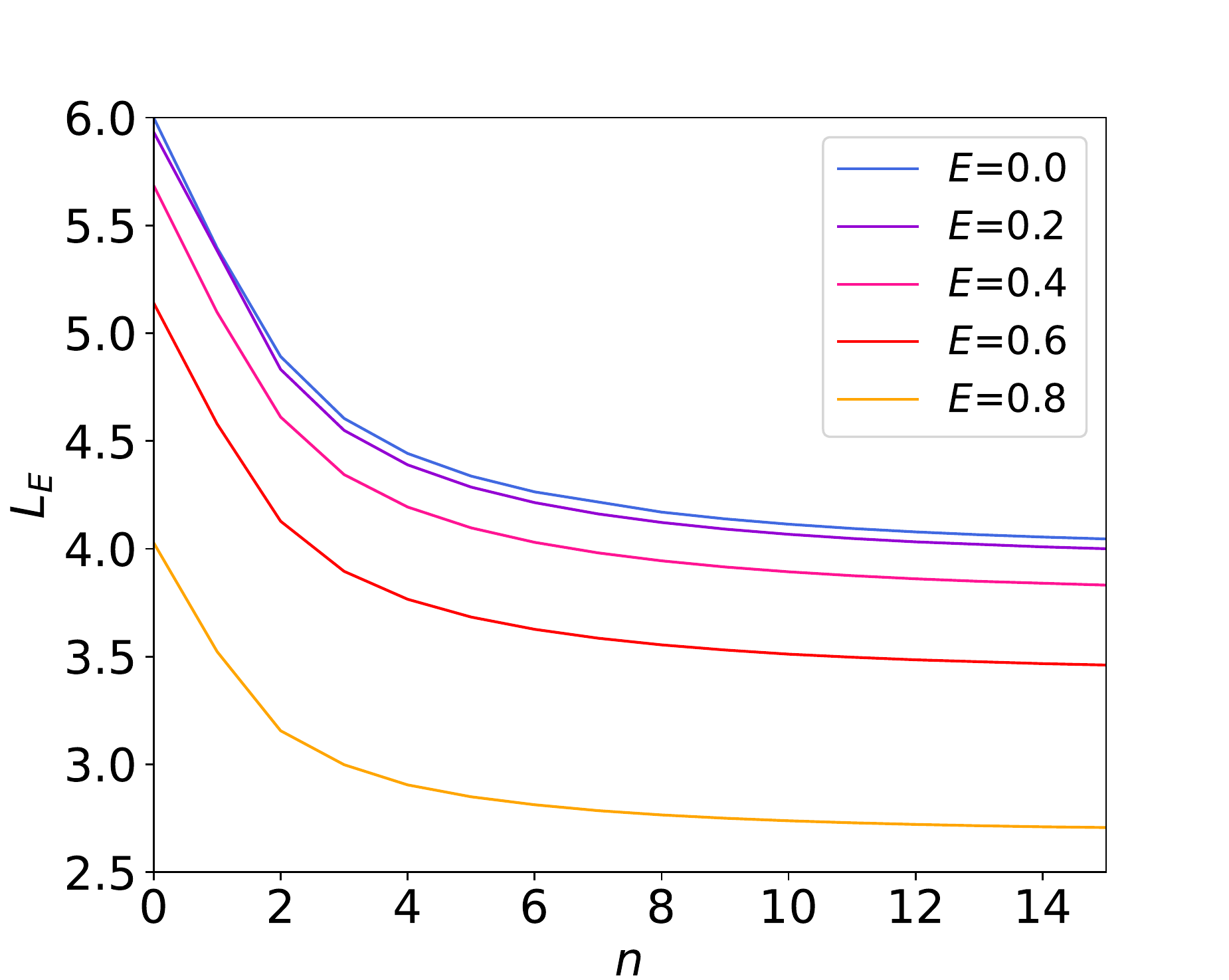}
	\includegraphics[scale=0.30]{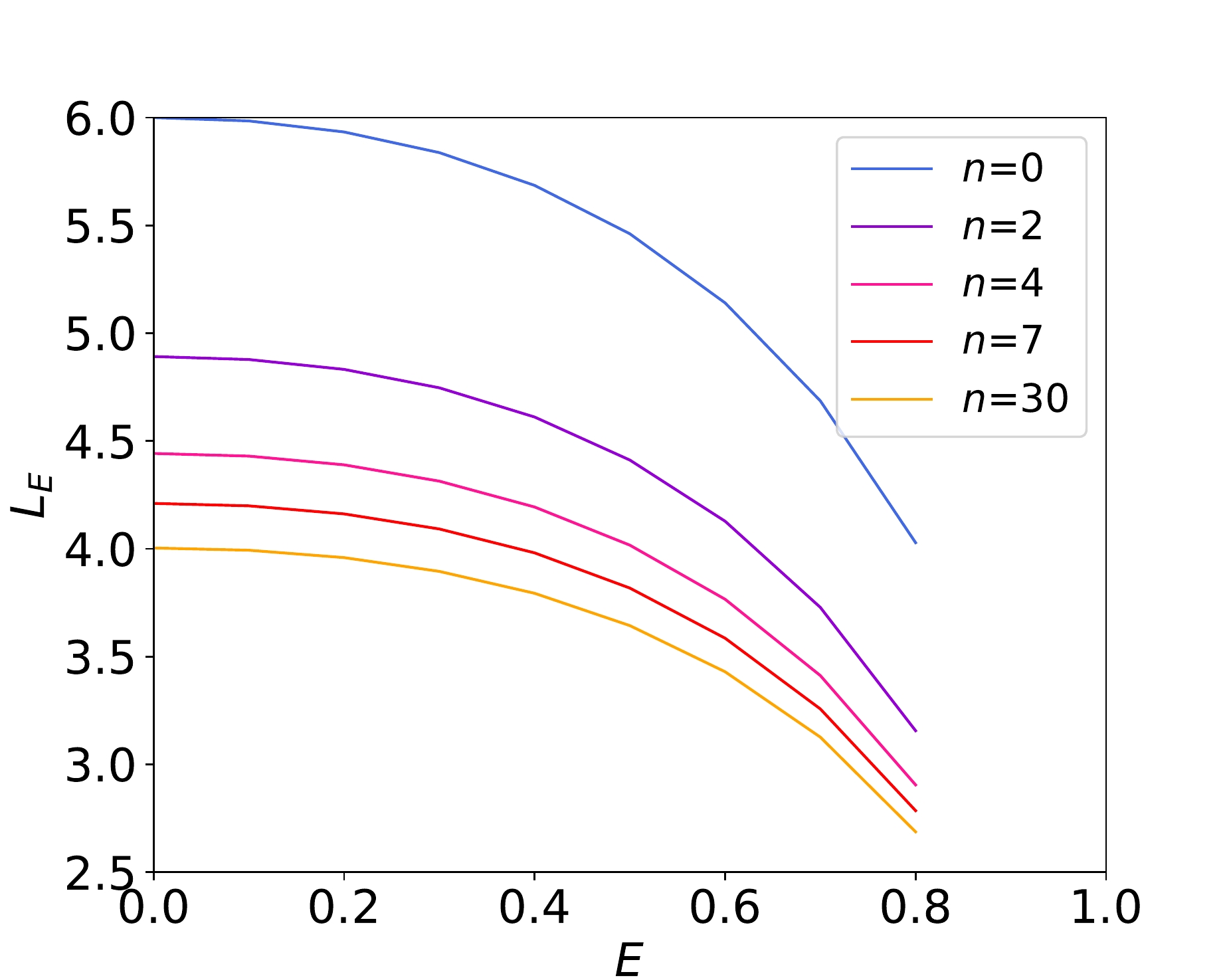}
	\includegraphics[scale=0.30]{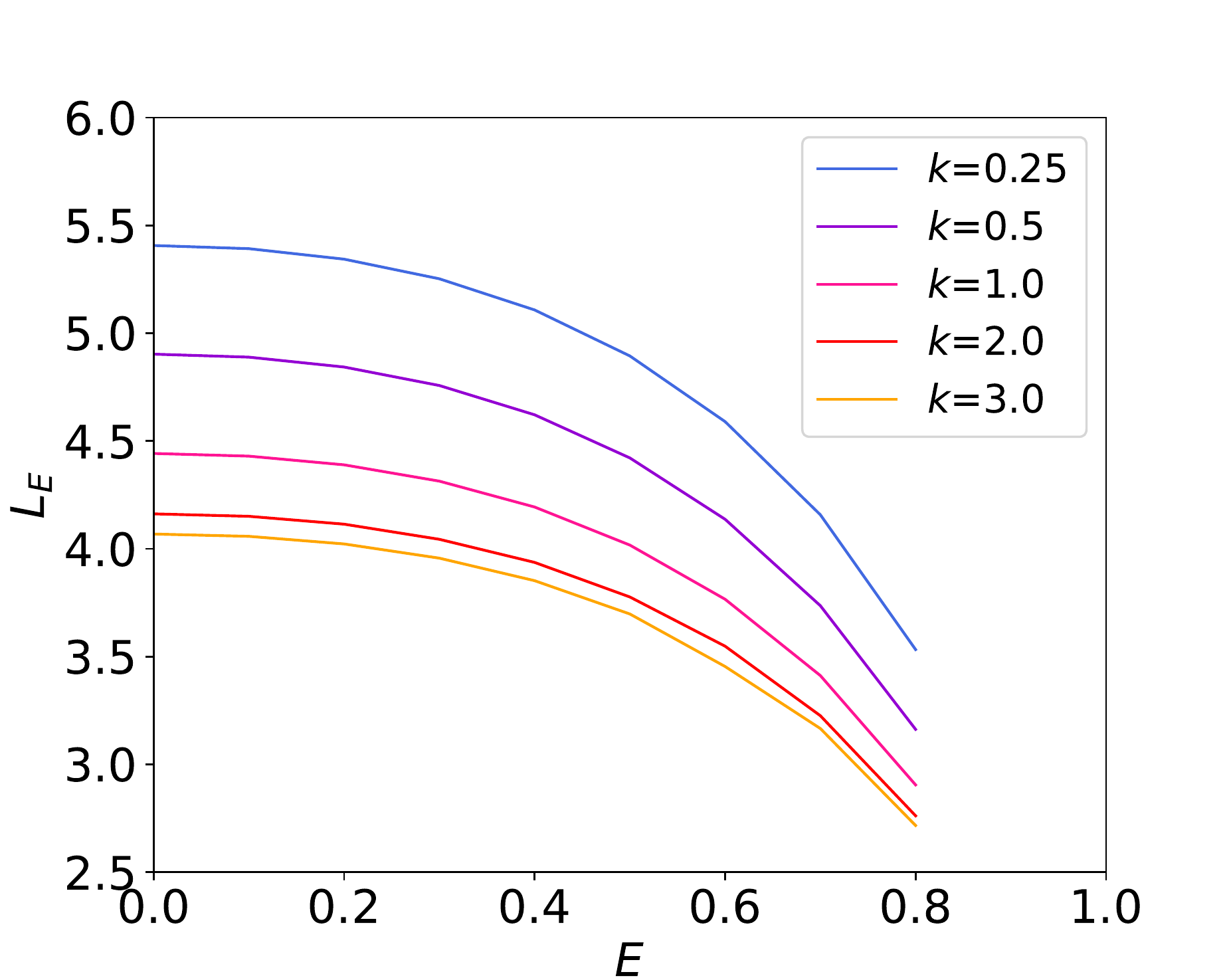}
	\caption{Local extraction of energy from noisy quantum battery. We plot the maximum amount of energy that can be extracted from the noisy battery which was initially described by a pure state, $\ket{\xi_E}$, with fixed entanglement $E$. All other considerations are same as Fig. \ref{fig2}.}
	\label{fig3}
\end{figure*}

We begin by observing how much energy can be extracted from noiseless quantum batteries, being in a pure locally passive state of fixed entanglement, using global unitary operations. Let us first construct the locally passive states, $\ket{\xi^l_E}$, having fixed entanglement, $E$.

Any pure state can be written in Schmidt decomposition. Hence we can write the locally passive state, $\ket{\xi^l_E}$, in the following way
\begin{equation*}
    \ket{\xi^l_E}=\sqrt{\lambda_1} \ket{\phi_1}\ket{\psi_1}+\sqrt{\lambda_2} \ket{\phi_2}\ket{\psi_2},
\end{equation*}
were, $\lambda_i$s denote the eigenvalues of the subsytem $\text{Tr}_{1}(\ket{\xi^l_E}\bra{\xi^l_E})$ or $\text{Tr}_{2}(\ket{\xi^l_E}\bra{\xi^l_E})$, and $\ket{\phi_i}$ and $\ket{\psi_i}$ are the corresponding eigenvectors of $\text{Tr}_{2}(\ket{\xi^l_E}\bra{\xi^l_E})$ and $\text{Tr}_{1}(\ket{\xi^l_E}\bra{\xi^l_E})$, respectively.
Since we are restricting the entanglement between the two qubits to be $E$, the eigenvalues can be written in terms of $E$ in the following way
\begin{equation*}
    \lambda_i=\frac{1}{2}\left(1\pm \sqrt{2^{E+1}-2^{2E}}\right).
\end{equation*}
 Since $\ket{\xi^l_E}$ is locally passive, its sub-systems, described by $\text{Tr}_{2}(\ket{\xi^l_E}\bra{\xi^l_E})$ and $\text{Tr}_{1}(\ket{\xi^l_E}\bra{\xi^l_E})$, have to commute with $\sigma_1^z$ and $\sigma_2^z$, respectively, and also the eigenvalues of $\text{Tr}_{2/1}(\ket{\xi^l_E}\bra{\xi^l_E})$ should be in an increasing order. Hence the locally passive state, $\ket{\xi_E^l}$, having fixed entanglement, $E$, has the following form
\begin{equation*}
    \ket{\xi_E^l}=\frac{1}{2}\left(1- \sqrt{2^{E+1}-2^{2E}}\right)\ket{00}+\frac{1}{2}\left(1+ \sqrt{2^{E+1}-2^{2E}}\right)\ket{11}.
\end{equation*}
If we assume that initially the quantum battery is in the locally passive state, $\ket{\xi^l_E}$, then the maximum amount of energy that can be extracted globally from the battery can be obtained using Eq. \eqref{eq3} and is given by \cite{ref20}
\begin{equation*}
    G_E^p(0)=(e_1+e_2)\left(1-\sqrt{2^{E+1}-2^{2E}}\right).
\end{equation*}
Within the time $n\delta t$, the second qubit of the battery will interact with $n$ distinct spins of the bath. The extractable energy, $G_E^p(n)$, from the noisy quantum battery, after time $n \delta t$, will depend on the entanglement of the initial state as well as the number of times it has interacted, i.e $n$. In the left panel of Fig. \ref{fig1}, we plot the nature of the $G_E^p(n)$ as a function of $n$ for different values of $E$. Though $n$ takes integer values, to realize the behavior of the envelope, we have joined the discrete points taken in an interval of $\delta t$ with lines. The behavior of the quantity in between the intervals, $\delta t$, will be analyzed in Sec. \ref{sec5}. In the middle and right panel of the figure, we plot $G_E^p(n)$ as a function of $E$, for different values of $n$ and $k$, respectively. To plot the right panel, we kept the value of $n$ fixed at four. From the figure, we can conclude that the nature of the qualitative dependence of $G_E^p(n)$ on $E$ does not change with $n$. Quantitatively, the amount of extractable work, $G_E^p(n)$ decreases with the increase in $n$. This is intuitively satisfactory, since during each interaction of the battery with the qubit the battery looses some of its energy to the bath. After a threshold value of $n$ (around $n\approx 5$) the decrease in extractable energy becomes negligible, i.e., $G_E^p(n)$ becomes almost constant about $n$. In the right panel of Fig. \ref{fig1}, we kept $n$ fixed at $n=4$. It is evident from the right panel that the amount of extractable energy also depends on the coupling: as we increase the interaction, $k$, more and more energy gets lost to the bath. 

The curves of the middle and right panels of Fig. \ref{fig1} follow the behavior of the function $3c\left(a-\sqrt{2^{E+1}-2^{2E}}\right)$, where the values of the parameters, $a$ and $c$, depend on $n$ (for middle panel) and $k$ (for right panel). For example, when we fit the red (pink) curve shown in the middle panel (right panel) for $n=7$, ($k=1.0$) the obtained parameter values are $c=0.54$ ($c=0.68$) and $a=1.0$ ($a$=1.0). The error in the fitted function and the 95\% confidence errors are of the order of $10^{-6}$. We have used the least square fitting method to obtain the best possible fitting.

In the next subsection, instead of being restricted to only locally passive states, we will consider the whole set of pure states and explore the energy extraction properties there by considering the same situation.

\subsection{Extraction of work from pure battery states}

Let us now consider the set of all pure states, $\{\ket{\xi_E}\}_E$, which have a fixed entanglement, $E$. If the amount of extractable energy from a pure state is maximized over the set of states $\{\ket{\xi_E}\}_E$, the resulting amount will be of the form \cite{ref20}
\begin{equation*}
   G_E(0)=(e_1+e_2)\left(1+\sqrt{2^{E+1}-2^{2E}}\right).
\end{equation*}
But if we consider the interaction between the battery and the bath, $\mathcal{B}$, the amount of extractable work will change depending on the number of spins it interacted with, i.e., $n$, and the coupling constant, $k$. Let us consider the battery initially prepared in a pure state, $\ket{\xi_E}$, having fixed entanglement, $E$. The maximum amount of extractable energy from the battery, $G_E(n)$, maximized over all possible initial pure states with fixed entanglement $E$, after $n$ consecutive interactions of the second qubit of the battery with the bath, is shown in Fig. \ref{fig2}. In the left panel we plot $G_E(n)$ as a function of $n$ and in the middle and right panels we plot the same but as a function of $E$. The difference between the middle and right panels is that in the middle panel we have shown the behavior of the function for different values of $n$, whereas in the right panel it is depicted for different $k$ values and a fixed $n$ value, i.e., $n=4$.
It is apparent from Fig. \ref{fig2} that the extractable work decreases with increase in entanglement as well as the number of times the state interacted with the bath. Moreover, for a given value of $E$, and much larger values of $n$, $G_E(n)$ becomes almost constant. Here also $G_E(n)$ decreases with $k$, i.e., the coupling between the second qubit and the spin. The curves, plotted in the middle panel of Fig. \ref{fig2}, can be well fitted by the function $3c\left(1+\sqrt{2^{E+1}-2^{2E}}\right)$ for $n<4$. In the region $n\geq 4$, the curves can be fitted using a three-parameter (say $a$, $b$ and $c$) function $3c\left(1+\sqrt{2^{E+1}-2^{2E}}\right)+be^{aE}$.
An appropriate fitting function for the curves in the right panel is $3p(1+\sqrt{2^{E+1}-2^{2E}})+qe^{rE^3}$, where $p$, $q$, and $r$ are fitting parameters.  As an example, if we consider the curves corresponding to $n=7$ and $k=1$,  depicted, respectively, in the middle and right panels, the corresponding values of the fitting parameters will be $a=0.029\pm 0.0083$, $b=-1.2\pm 0.034$, $c=0.89\pm 0.0054$, $p=0.91\pm 0.079$, $q=-1.0\pm 0.48$, and $r=-0.065\pm 0.12$. The errors in the least square fittings are 0.0034 and 0.0027. The number appearing after $\pm$ sign represents the 95\% confidence error.

\subsection{Local work extraction from noisy quantum battery}
In this sub-section, we want to explore the effect of the interaction with the bath on the amount of extractable work using local operations. Again we consider the same set of pure states, $\{\ket{\xi}_E\}_E$, having entanglement $E$. But in this case, instead of maximizing the globally extractable energy, we maximize the amount of energy that can be extracted using local operations only. The maximum locally extractable energy from the states $\{\ket{\xi}_E\}$ is given by \cite{ref20}
\begin{equation*}
    L_E(0)=2(e_1+e_2)\sqrt{2^{E+1}-2^{2E}}.
\end{equation*}

Let the battery initially be in one of the pure states $\{\ket{\xi}_E\}_E$. The behavior of $L_E(n)$, that is the maximum locally extractable energy from the battery, after $n$ number of interactions with the bath's spins, is shown in Fig. \ref{fig3}. The maximization is performed over the set $\{\ket{\xi}_E\}_E$. The left panel represents the behavior of $L_E(n)$ whereas the other two panels shows the behavior of the same but as a function of $E$. Different values of $n$ and $k$ are considered to examine the nature of $L_E(n)$ in the middle and right panels, respectively, as mentioned in the legend. We keep the value of $n$ constant at four for the curves shown in the right panel. We see the variation of $L_E(n)$ with $E$, for a fixed $n$, does not change qualitatively, though the overall value decrease with increase in $n$. But for $n>5$, $L_E(n)$ becomes almost constant about $n$, for a fixed entanglement $E$ and coupling constant $k$. The rightmost panel of Fig. \ref{fig3} shows that $L_E(n)$ also decreases with increase in the coupling constant, $k$. We can fit the plotted curves, shown in the middle and right panels of Fig. \ref{fig3}, using the function $6c\left(\sqrt{2^{E+1}-2^{2E}}\right)+be^{ax}$, where $a$, $b$, and $c$ are fitting parameters dependent on the values of $n$ (for middle panel) or $k$ (for right panel). If we fit the $n=7$ and $k=1.0$ curve presented, respectively, in the middle and right panels, the values of the fitting parameters will be $a=0.12\pm 0.043$, $b=-0.10+0.0092$, and $c=0.72\pm 0.0014$, for the former and $a=0.093\pm 0.032$, $b=-0.20\pm 0.014$, and $c=0.77\pm 0.0022$, for the later. The least square errors present in these fittings are 0.0010 (corresponding to the $n=7$ curve of the middle panel) and 0.0016 (corresponding to the $k=1$ curve of the right panel).

\section{Effect of non-Markovianity}
\label{sec5}
\begin{figure}[h!]
	\includegraphics[scale=0.45]{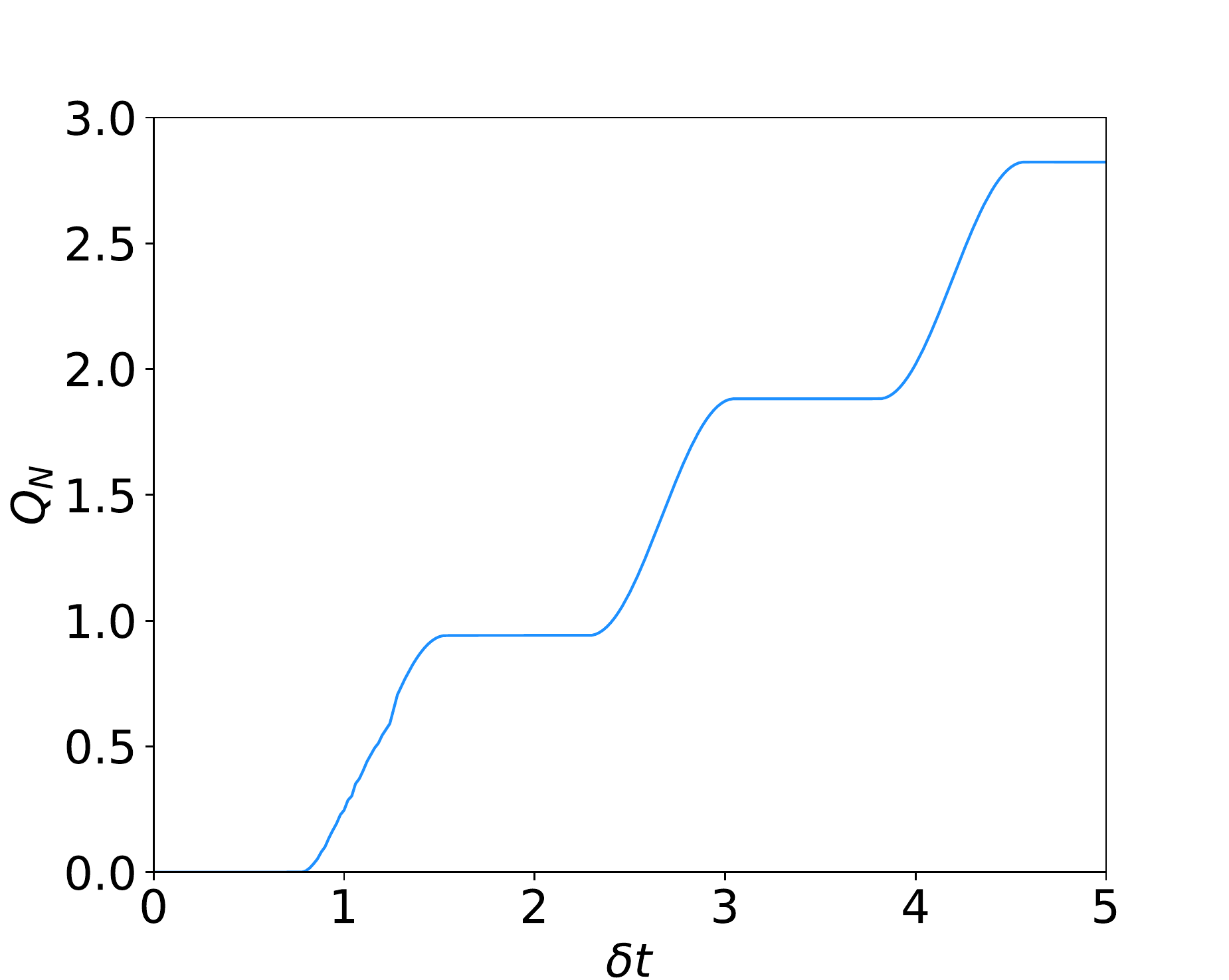}
	\caption{Amount of non-Markovianity present in the interaction with a single spin. After quantifying the non-Markovianity using the measure expressed in Eq. \eqref{QN}, we plot $Q_N$ as a function of the time duration, $\delta t$, of interaction with each single spin. The dimensionless vertical and horizontal axes represent the values of $Q_N$ and $\delta t$, respectively.}
	\label{fig5}
\end{figure}
\begin{figure*}
\centering
	\includegraphics[scale=0.30]{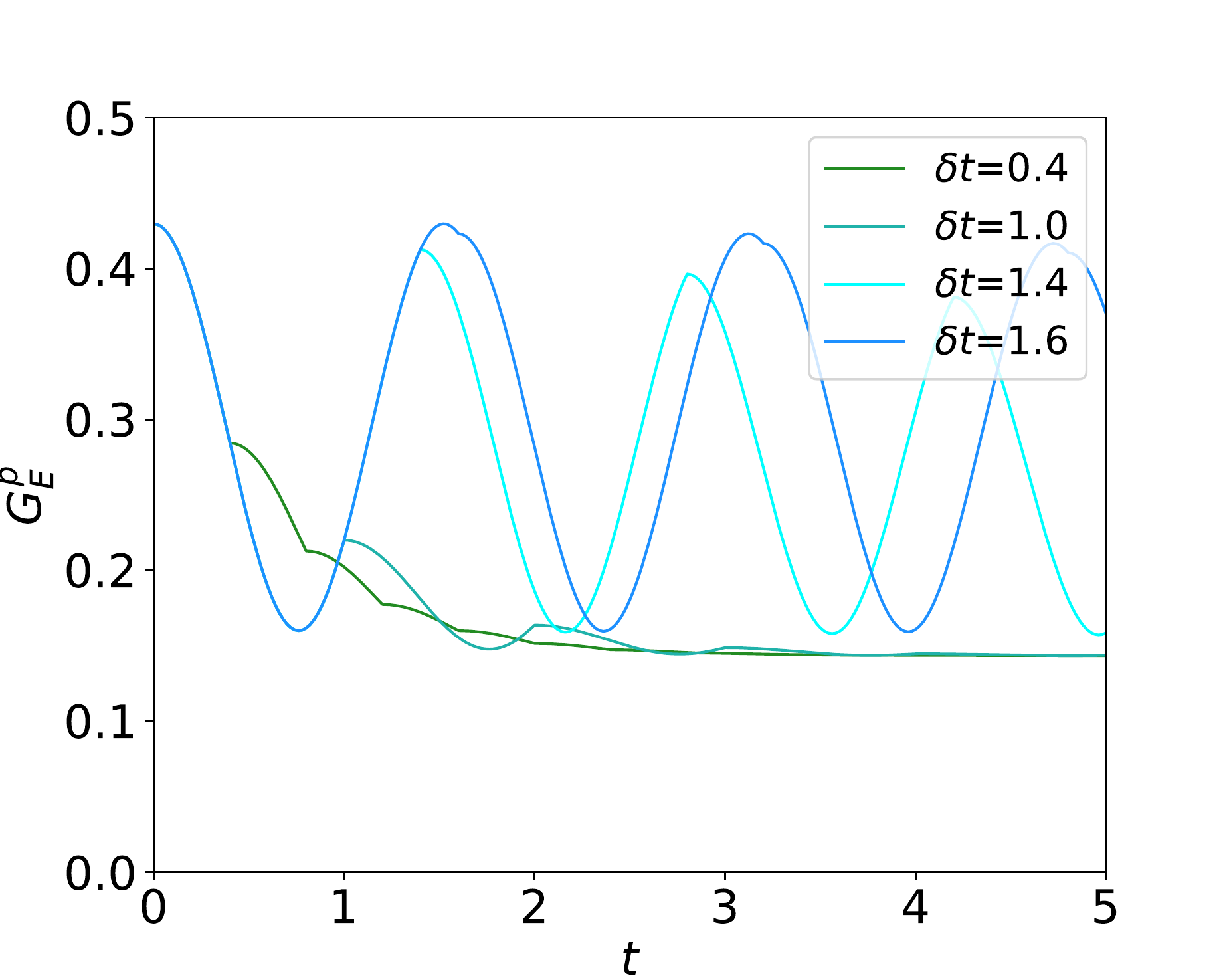}
	\includegraphics[scale=0.30]{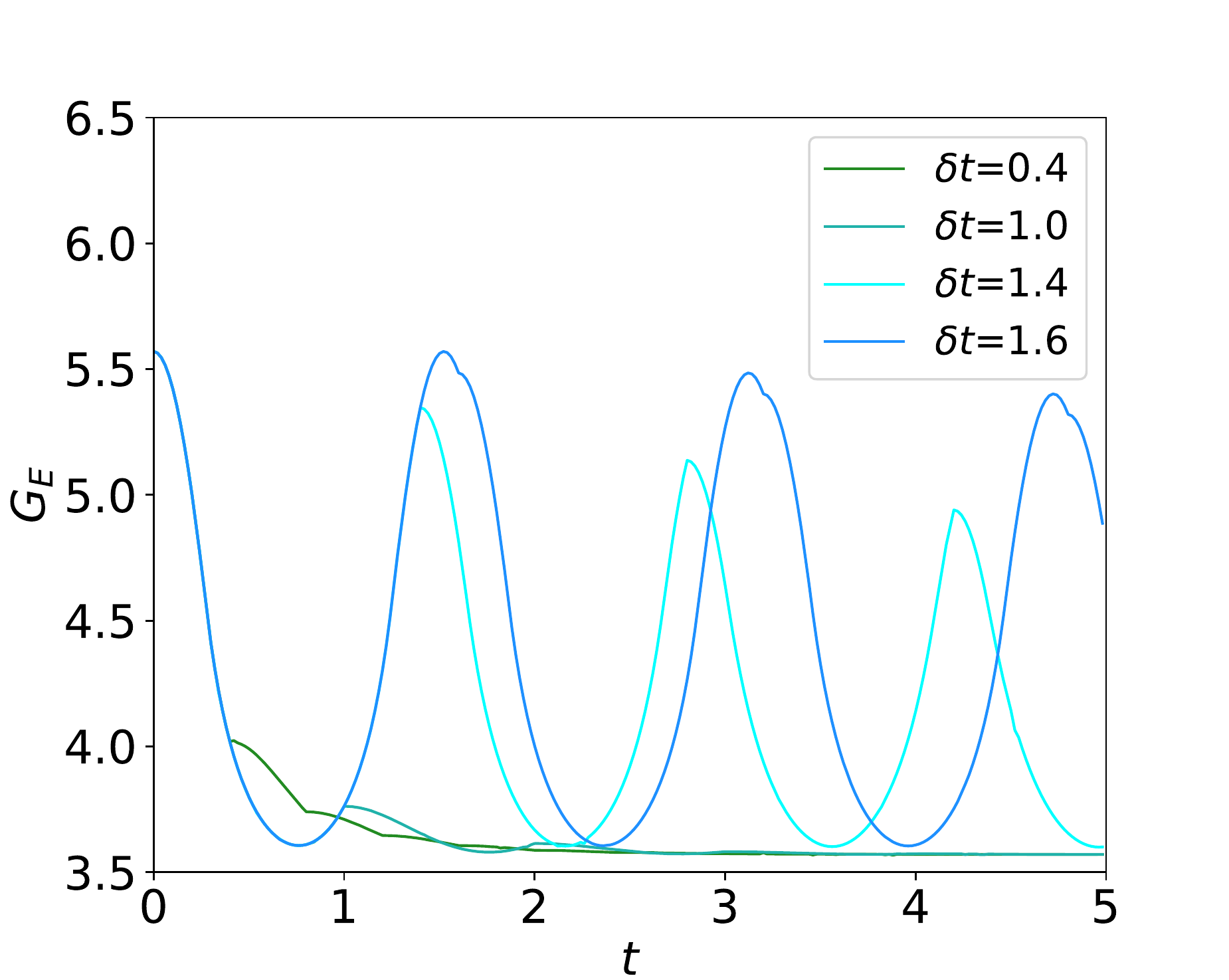}
	\includegraphics[scale=0.30]{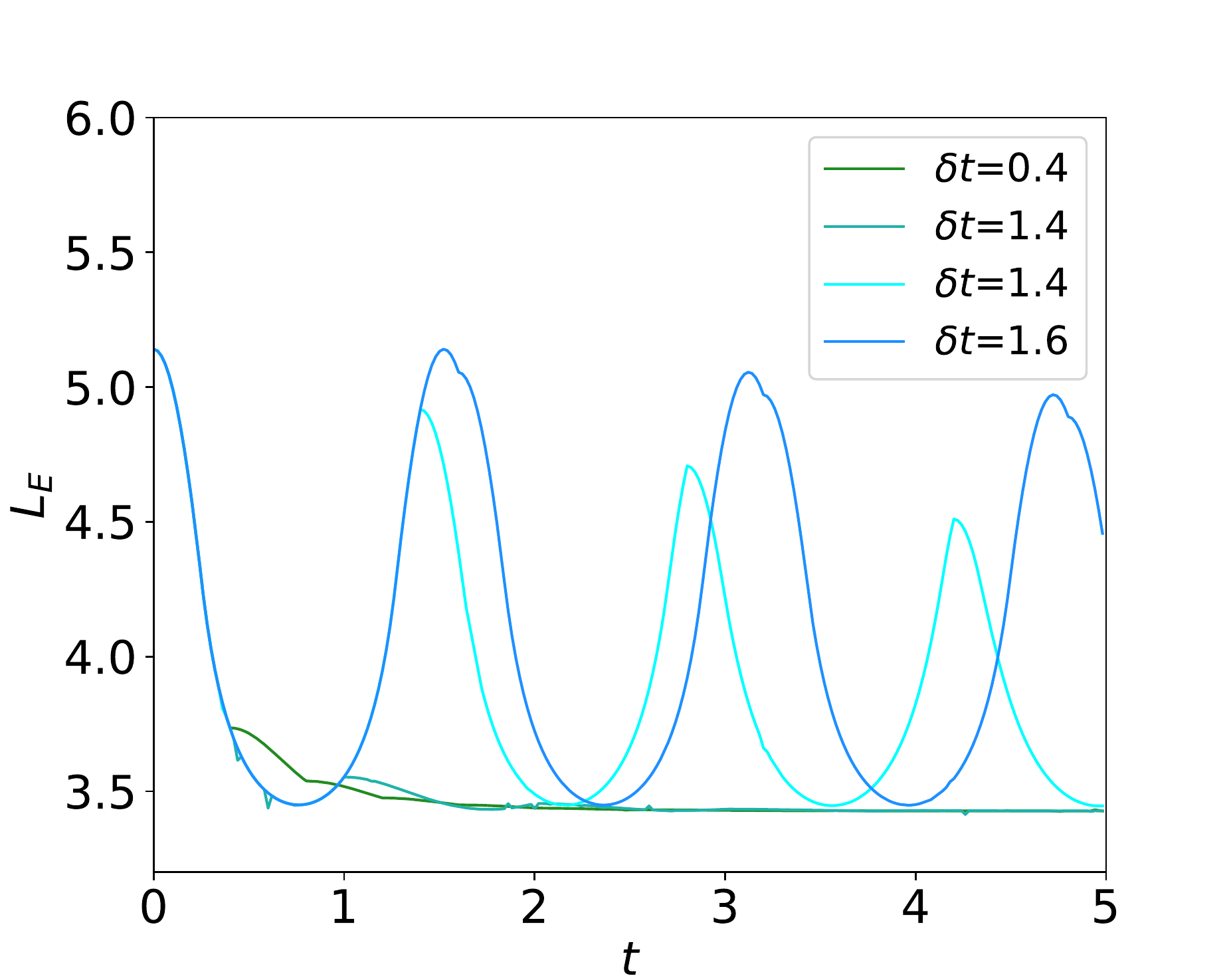}
	\caption{Extractable energy from the interacting quantum battery. We present the behavior of the maximum amount of energy that can be extracted from a battery, initially being in a locally passive state (left panel) or any pure state (right panel), with a fixed entanglement, $E=0.6$, after its interaction with the bath for $t$ amounts of time. In the middle panel, we show the maximum locally extractable energy from battery initially prepared in a pure state with the same entanglement, $E=0.6$, as a function of $t$. The vertical axes represent $G_E^p$ (left panel), $G_E$ (middle panel), and $L_E$ (right panel) in units of $e$. The dimensionless horizontal axes represent the time of total interaction, i.e., $t$. In all of the three plots, we considered a set of values of the interaction time with each spin, $\delta t$, given by $\delta t=$0.4 (green line), 1.2 (sea-green line), 1.4 (cyan line), and 1.6 (blue line).}
	\label{fig4}
\end{figure*} 

In the considered model, each spin of the bath acts for time duration $\delta t$, and after the interaction, a new spin arrives to interact with a new initial state that has no memory of the previous interactions. Up until now, we have examined the battery after each of these intervals, $\delta t$, not in between the intervals. In this section, we want to explore the nature of the extractable energy within these intervals and analyze the impact of non-Markovianity present in the interaction.

As we have mentioned, after each interval, a new spin comes to interact with the qubit. As a result, the effect of non-Markovianity disappears immediately after each time duration $\delta t$. But non-Markovianity can come into play within the intervals, and hence it is straight-forward that it will depend on the length of the time interval, $\delta t$. Before going into the details of extractable energy, let us first try to quantify the non-Markovianity. To assess non-Markovianity, we employ the Bruner et al.~\cite{NM} measure. In the presence of Markovian interactions, systems slowly lose their identity over time. Thus two systems, initially having two distinct states, say $\rho_1(0)$ and $\rho_2(0)$, acting on the same Hilbert space, if gets affected by a Markovian interaction, $M$, will become less distinguishable. Thus the distance, $D(\rho_1(t,M),\rho_2(t,M))$, between the states, $\rho_1(t,M)$ and $\rho_2(t,M)$,  monotonically decreases with time  of interaction, $t$. Conversely, increase in $D(\rho_1(t,M),\rho_2(t,M))$ with time is an evidence of non-Markovianity. The non-Markovianity measure, $Q_N$, of a channel, $M$,  presented in Ref.~\cite{NM}, is defined based on this concept and is given by
\begin{equation}
Q_N(M)=\max_{\rho_1(0),\rho_2(0)}\int_{\Lambda>0} \Lambda(t,M)dt,
\label{QN}
\end{equation}
where $\Lambda(t,M)=\frac{d}{dt}D(\rho_1(t,M),\rho_2(t,M))$.

In this work, we have considered the distance measure to be trace distance, i.e., $D(\rho_1(t,M),\rho_2(t,M))=\frac{1}{2}\text{Tr}|\rho_1(t,M)-\rho_2(t,M)|$. We have used the notation $|A|=\sqrt{A^\dagger A}$. For the quantification of non-Markovianity, we have restricted the coupling constant at $k=1$. To simplify numerical complexity, we have maximized over only pure states.  In Fig. \ref{fig5}, we plot the non-Markovianity, $Q_N$, by considering the interaction of the battery with a single spin of the bath, as a function of the time duration of that interaction, $\delta t$. We see up to $\delta t\approx 0.78$, the interaction is Markovian. That is because within this small amount of time, the system only loses information to the environment and is unable to collect any from it. But when we increase $\delta t$ further the system starts to gain information back from the bath and thus $Q_N$ becomes non-zero.

To experience the effect of non-Markovianity, in Fig. \ref{fig4}, we plot $G_E^p$ (left panel), $G_E$ (middle panel), and $L_E$ (right panel ) against the total time of interaction, $t$. We kept the initial shared entanglement fixed at $E=0.6$ ebits and considered four different values of $\delta t$. At $\delta t=0.4$, the interaction is Markovian and the corresponding extractable energies also have different behavior that is $G_E^p(n)$, $G_E(n)$, and $L_E(n)$ are non-increasing with time. But if we increase $\delta t$ to 1.0 or more, within the time interval, $\delta t$, at first the extractable energy decreases, but after a point it starts to increase. This can be explained as follows: when the environment interacts with the battery, it absorbs energy from the system at first, but after a while, the spin begins to exhale the energy back into the system, resulting in an increase in the amount of extractable energy. Thus, if we extract the energy within an appropriate range of time, we can get a non-Markovian improvement in the accessible energy compared with the Markovian scenario. After each time interval, $\delta t$, a new spin comes to interact and again starts to take away energy. As a result, the kinks in the curves are visible after each time interval, $\delta t$.
\section{Conclusion}
\label{sec6}
Though quantum batteries are initially introduced as an isolated system, in practical scenarios, it is impossible to entirely isolate the battery from its environment. The device's unavoidable or unrecognized interactions with the environment can have an impact on its operational efficiency.

Here we consider a two-qubit quantum battery, one of which is involved in successive interactions with the spins contained in a bath. We consider the situation where a single spin interacts at a time for a fixed amount of time, after which a new spin with no knowledge of the prior interactions comes to interact for the same amount of time. This process goes on. 

The battery is initially prepared in a locally passive state with fixed entanglement. We maximize the globally extractable work from the battery over its possible initial states and observe its behavior as a function of the entanglement shared between the constituents of the batteries or the number of successive interactions with the bath. The extractable work appears to decrease with the number of interactions before saturating at a constant value.

In the next setup, we remove the restriction on the initial state of the battery being locally passive and instead consider it to be any pure state but still with a fixed entanglement. By considering such a set of initial states, we maximize the amount of work extractable using global and local operations, considering its interaction with the bath. Again, both the locally and globally extractable work, though having the qualitatively same behavior with entanglement, initially decreases with the number of spins it interacts with before finally reaching an almost constant value.

If we examine the interaction after each time interval of the single-spin interaction, the interaction will seem to be non-Markovian. But if we look at the behavior in between the time intervals, the interaction may appear non-Markovian. We quantify the amount of this non-Markovianity arriving in between the interactions as a function of the time duration of the interaction with each spin, which is the same for all spins. We also explore how the non-Markovianity can improve the amount of extractable work, be it local or global, in between the interactions.

\end{document}